\documentclass{nature}
\usepackage{graphicx}
\usepackage{amssymb}

\title{Ultra High Energy Cosmic Ray Acceleration in \\ Engine-driven Relativistic Supernovae}

\author{S. Chakraborti$^{1,2}$, A. Ray$^{1,2}$, A.M. Soderberg$^2$, A. Loeb$^2$ \& P. Chandra$^3$}

\begin{document}

\maketitle

\begin{affiliations}
 \item Tata Institute of Fundamental Research, 1 Homi Bhabha Road, Mumbai 400 005, India
 \item Harvard-Smithsonian Center for Astrophysics, 60 Garden Street, MS-51, Cambridge, 
 \\Massachusetts 02138, USA. 
 \item Royal Military College of Canada, Kingston, Ontario, Canada K7K 7B4
\end{affiliations}

\begin{abstract}
The origin of the highest energy cosmic rays remains an enigma.
They offer a window to new physics, including tests of physical laws
relevant to their propagation and interactions, at energies unattainable
by terrestrial accelerators. They must be accelerated locally, as otherwise
background radiations would severely suppress the flux of protons and nuclei,
at energies above the Greisen-Zatsepin-Kuzmin (GZK)
limit
($\sim60$EeV=$6 \times 10^{19}$eV).
Nearby Gamma Ray Bursts (GRBs),
Hypernovae,
Active Galactic Nuclei (AGNs)
and their flares,
have all been suggested and debated as possible sources.
A local sub-population of type Ibc supernovae (SNe) with mildly relativistic outflows
have been detected as sub-energetic GRBs
or X-Ray Flashes (XRFs)
and recently as radio afterglows without detected GRB counterparts.
We measure the size-magnetic field evolution, baryon loading and energetics, using the
observed radio spectra of SN 2009bb. We place such engine-driven SNe above the Hillas
line
and establish that they can readily explain the post-GZK UHECRs.
\end{abstract}

The highest energy cosmic rays pack such a large amount of energy
and have such a low flux\cite{2007Sci...318..938T} that direct
detection by satellite-borne instruments is infeasible, allowing
them to be detected\cite{1963PhRvL..10..146L} only by cosmic ray air
showers\cite{1937RSPSA.159..432B}.
UHECRs beyond the GZK limit\cite{1966PhRvL..16..748G,1966JETPL...4...78Z}
have been invoked
to propose tests of known physical laws and symmetries\cite{1999PhRvD..59k6008C}.
Understanding their origin is a crucial step in using them as probes of new physics.
But, the sources of the highest energy cosmic rays pose an intriguing
problem, since the magnetic rigidity of these particles are such that
the magnetic fields in our
galaxy are neither strong enough to contain them nor bend them
sufficiently\cite{1963PhRvL..10..146L}. Yet, among the UHECRs which have
been detected until now, no concentration have been found towards the
Milky Way. Hence, it is anticipated that their sources are extragalactic.
However, UHECR protons with energies above 60 EeV can interact with a
significant fraction of CMB photons via the $\Delta$ resonance.
The cross section of this interaction is such that only those extragalactic cosmic
ray sources locally (within 200 Mpc of the Earth) can contribute significantly to
the flux of UHECRs above the so called GZK
limit\cite{1966PhRvL..16..748G,1966JETPL...4...78Z}. At least 61 sources are
required by observations of UHECRs until now, to provide cosmic rays
with energies above the GZK limit\cite{2007Sci...318..938T}.
Since particles of such high energy could not have traveled to the Earth from
cosmological distances, unless Lorentz invariance breaks down at these
energies\cite{1999PhRvD..59k6008C}, they encourage the search for potential
cosmic ray accelerators in the local Universe.
AGNs have been considered\cite{2007Sci...318..938T} as UHECR sources.
But, most of them are not luminous enough\cite{2009ApJ...693..329F},
leaving proposed very intense, short
duration AGN flares, which are yet to be observed, as possible
sources\cite{2009ApJ...693..329F}.
Classical Gamma Ray Bursts (GRBs) are also considered as possible
sources\cite{1995PhRvL..75..386W,1995ApJ...449L..37M}, but most of them occur
beyond the GZK horizon\cite{2006A&A...447..897J} and cannot contribute
significantly to the local flux beyond the GZK limit\cite{2009ApJ...693..329F}.
Hypernovae have also been suggested as sources where particles
are boosted to successively higher energies in an ejecta profile
extending upto mildly relativistic
velocities\cite{2007PhRvD..76h3009W,2008ApJ...673..928B}. However,
they require excessive explosion energy and fail to reproduce the flat
injection spectrum of UHECRs
(see Suppl.~Info
for discussion of some of the proposed sources).

Soon after their suggestion that SNe come from collapse of a normal
star to a neutron star, Baade and Zwicky went on to suggest that SNe
may be the source of cosmic rays as well\cite{1934PNAS...20..259B}.
Since then, SNe and SN remnants have been studied as sources of
high energy cosmic rays.
However, ordinary SNe and their remnants can
not produce UHECRs due to two fundamental limitations. Firstly,
they well lie below the line representing the
combination of size and magnetic field required to confine and accelerate Iron
nuclei with energies
of 60 EeV, in the so called Hillas diagram\cite{1984ARA&A..22..425H} (Figure 1).
The second, even more restrictive, condition obviously not fulfilled
by ordinary SNe is because ordinary SNe have $\beta/\Gamma\sim0.05$
($\beta\equiv v/c$ and $\Gamma\equiv 1/\sqrt{1-\beta^2}$, where $v$ is the speed
of the blastwave and $c$ is the speed of light in vacuum) 
which restricts the highest energy cosmic rays accelerated in ordinary SNe
to well below the GZK limit.

Until recently, SNe with relativistic ejecta have been
spotted exclusively through Long GRBs associated with them like
GRB 980425\cite{1998Natur.395..663K} or its twin GRB 031203.
The discovery of XRF 060218\cite{2006Natur.442.1014S} associated with SN 2006aj
showed that mildly relativistic SNe are hundred times less energetic but thousand
times more common (in their isotropic equivalent rate, which is relevant for UHECRs
actually reaching the observer) than classical
GRBs\cite{2006Natur.442.1014S}. Radio follow up of
SNe Ibc have now discovered the presence of an
engine driven outflow from SN 2009bb\cite{2010Natur.463..513S}, without
a detected GRB.
The mildly relativistic SNe, detected either using XRFs
or radio afterglows, a subset of SNe Ibc are far more abundant
at low redshifts required for the
UHECR sources, than the classical GRBs. Moreover, given their mildly  relativistic
nature, they have the most favorable combination of $\beta/\Gamma\sim1$, unlike
both non-relativistic SNe and ultra-relativistic classical Long GRBs.

In order to derive the highest energy upto which these relativistic SNe
can accelerate cosmic rays, we have to determine the
evolution of the size and the magnetic field in the blast-wave.
It has been demonstrated that a Synchrotron Self Absorption (SSA) model fits the
initial radio spectrum of SN 2009bb rather well,\cite{2010Natur.463..513S}
with a low frequency
turnover defining the spectral peak shifting to lower frequency with
time, characteristic of the expansion of the shocked region that
powers the radio emission. This allows us to measure the evolution of
the radii and magnetic fields from VLA and GMRT data (see Suppl.~Info) at 5 epochs,
plotted on the Hillas diagram (Figure 1). This clearly
demonstrates that SN 2009bb and XRF 060218 can both confine UHECRs and accelerate
them to highest energies seen experimentally.
At the time of the earliest radio observations\cite{2010Natur.463..513S}
with its fortunate combination of $\beta/\Gamma\sim1$, SN 2009bb
could have accelerated nuclei of atomic number $Z$ to an
energy of $\sim6.5\times Z$ EeV. For example, the source could have
accelerated protons, Neon, and Iron nuclei to 6.4, 64 and 166
EeV respectively. In this scheme, the highest energy particles are
likely to be nuclei heavier than protons, consistent with the latest
results indicating an increasing average rest mass of primary UHECRs
with energy\cite{2010PhRvL.104i1101A}. Therefore, our results
support the claimed preference of heavier UHECRs at the highest energies
of the Auger collaboration, although this claim is disputed by another
experiment\cite{2010PhRvL.104p1101A}.

To estimate whether there are enough relativistic SNe to explain the
target objects associated with the $\sim60$ detected UHECRs,
we require the rate of such transients. SNe
Ibc occur at a rate\cite{1999A&A...351..459C,2004ApJ...613..189D}
of $\sim1.7\times 10^4$ Gpc$^{-3}$ yr$^{-1}$. The
fraction of Ibc which have relativistic outflows is still a somewhat uncertain
number, estimated\cite{2010Natur.463..513S} to be around $\sim0.7\%$.
Hence the rate of SN 2009bb-like mildly relativistic
SNe is $\sim1.2\times 10^{-7}$ Mpc$^{-3}$ yr$^{-1}$, which is
comparable to the rate of mildly relativistic SNe detected as
sub-energetic GRBs or XRFs of $\sim2.3\times 10^{-7}$ Mpc$^{-3}$ yr$^{-1}$.
This gives us $\sim4$ (or $0.5$) such objects
within a distance of 200 (or 100) Mpc every year.
Since SN 2009bb is still a unique object, only a systematic radio survey
can establish their cosmic rate and statistical properties (see Suppl.~Info). 
However, cosmic rays of different
energies have different travel delays due to deflections by magnetic
fields. For a conservative mean delay\cite{2009ApJ...693..329F} of
$\langle \tau_{delay} \rangle \approx 10^5$ yrs
we may receive cosmic rays from any of $4$ (or $0.5$) $\times10^5$ possible
sources at any point in time.
Given the situation, in which a direct association between a detected
UHECR and its source is unlikely\cite{2008JCAP...05..006K}, the literature
in the subject has focused on the
constraints\cite{1984ARA&A..22..425H, 2009JCAP...08..026W} placed on
plausible sources. We have shown in our work
that indeed this new class of objects satisfy all these constraints.

Nuclei are also subject to photo-disintegration
by interaction with Lorentz boosted CMB photons and can travel upto a distance
of $\sim 100$ Mpc (see Suppl.~Info), smaller than but comparable to the GZK
horizon. So, the local rate of mildly relativistic
SNe is high enough to provide enough ($\gg60$) independent sources of
cosmic rays with energies above the GZK limit.
The value of $\langle \tau_{delay} \rangle$ also implies
that it will not be possible to detect UHECRs from a known relativistic SN,
such as SN 2009bb, within human timescales. However, high
energy neutrinos from photo-hadron interaction at the acceleration site may
be a prime focus of future attempts at detecting these sources with neutrino
observatories like the IceCube (see Suppl.~Info).

The required energy injection rate per logarithmic interval in
UHECRs\cite{1995PhRvL..75..386W,2008AdSpR..41.2071B} is
$\Gamma_{inj}=(0.7-20)\times10^{44}$ erg Mpc$^{-3}$ yr$^{-1}$. Given the
volumetric rate of mildly relativistic SNe in the local universe,
if all the energy injected into UHECRs is provided by local mildly
relativistic SNe, then each of them has to put in around
$E_{SN}=(0.3-9)\times10^{51}$ ergs of energy, which is comparable to
the kinetic energy in even a normal SN and can easily be supplied
by a collapsar model\cite{1999ApJ...524..262M}. The minimum energy
in the relativistic outflow of SN 2009bb, required to explain the radio
emission alone, was found\cite{2010Natur.463..513S} to be $E_{eq}\approx10^{49}$
ergs. Moreover, the
mildly relativistic outflow of SN 2009bb has been undergoing
almost free expansion for $\sim1$ year. Our measurements of this
expansion allows us to show (see Suppl.~Info) that this
relativistic outflow, without a detected GRB, is
significantly baryon loaded
and the energy carried by the relativistic baryons is
$E_{Baryons}\gtrsim3.3\times10^{51}$ ergs.

If a relativistic outflow carries similar energies in protons, electrons
and magnetic fields, the radiative cooling of the electrons will
lead to an X-ray transient\cite{2009JCAP...08..026W}. In our model,
the acceleration occurs in the forward shock produced by the engine driven
relativistic ejecta, characterized by a single bulk Lorentz factor.
Protons and nuclei in such a collisionless shock show 
a flat spectrum of UHECRs\cite{1987PhR...154....1B},
consistent with the extragalactic component of the cosmic ray
spectrum\cite{2008AdSpR..41.2071B}. Our radio observations of SN 2009bb constrain
the energy carried by the radiating electrons\cite{2010Natur.463..513S}
and the energy of the relativistic baryons powering the almost free
expansion for $\sim1$ year until now (see Suppl.~Info).
These observations indicate a spectral index of $\approx1$ in the optically thin
part of the radio spectrum\cite{2010Natur.463..513S}.
This implies a power law distribution of relativistic electrons,
with an energy index between $p\approx(2-3)$, depending upon the relative positions
of the breaks in its spectra\cite{1999PhR...314..575P}.
The observed rate of relativistic SNe in the local universe
is consistent with the required rate of X-Ray\cite{2009JCAP...08..026W}
and radio (see Suppl.~Info) transients accompanying the
UHECR accelerators.

It has been found that the arrival direction of the Auger events correlate
well with the locations of nearby AGNs\cite{2007Sci...318..938T}, this
suggests that they come from either AGNs or objects with similar
spatial distribution as AGNs. Note that the HiRes events\cite{2008APh....30..175T}
do not show such a correlation.
Furthermore, UHECRs correlate well\cite{2008MNRAS.390L..88G}
with the locations of neutral hydrogen (HI) rich galaxies from
the HI Parkes All Sky Survey (HIPASS). Our proposal relies on the
acceleration of UHECRs in the mildly relativistic outflow from a subset of
SNe Ibc, for which we have determined the size and magnetic field
evolution using our radio observations, rather than hypothetical magnetars,
with as yet unknown magnetic fields, supposedly formed during sub-energetic
GRBs\cite{2008MNRAS.390L..88G} considered in that work.
SNe Ibc occur mostly in gas rich star forming spirals. In
particular the 21 cm fluxes of
NGC3278 (hosting
SN 2009bb) obtained from the HyperLeda database
amount to
$\sim1.9 \times 10^9 M_\odot$ of HI.
Hence, the observed correlation of UHECR arrival directions with HI selected
galaxies\cite{2008MNRAS.390L..88G} is consistent with our hypothesis.

In this letter we have shown that the newly established subset of nearby
SNe Ibc, with engine-driven mildly relativistic outflows detected as
sub-energetic GRBs, XRFs or solely via their strong radio emission,
can be a source of UHECRs with energies beyond the GZK limit. 
Our study demonstrates for the first time, a new class of objects, which
satisfy the constraints which any proposed accelerator of UHECRs
has to satisfy. If SN 2009bb
is characteristic of this newly discovered class, a radio survey to detect all
such events should be undertaken (see Suppl.~Info).
As an example, an all sky radio survey at $\nu=1$ GHz with a
sensitivity of 1 mJy and cadence of 2 months, can detect all such transient sources
which can accelerate Neon nuclei to 60 EeV, within 200 Mpc of the Earth
($\sim4$ per yr). Such a survey will also detect radio
emission from ordinary SNe and non-relativistic transients. However,
their faster rise to peak will easily separate out the relativistic SNe
for multi-frequency follow up.




\bibliography{uhecr}


\begin{addendum}
 \item SC thanks the lecturers at the 27th Winter School
in Theoretical Physics, Jerusalem. SC and AR acknowledge discussions with
Malcom Longair, Rohini Godbole and Shobo Bhattacharya. We thank the staff
of the GMRT who have
made some of the observations possible. GMRT is run by the National Centre for
Radio Astrophysics of the Tata Institute of Fundamental Research (TIFR).
The National Radio Astronomy Observatory is a facility of the National
Science Foundation operated under cooperative agreement by Associated
Universities, Inc. We acknowledge the usage of the HyperLeda database
(http://leda.univ-lyon1.fr).
AR would like to thank the Institute for Theory and Computation,
Harvard Smithsonian Center for Astrophysics for their hospitality.
At TIFR, which is celebrating the birth centenary
of its founder Homi J. Bhabha, this work is supported by Eleventh
Five Year Plan 11P-409.
 \item[Competing Interests] The authors declare that they have no
competing financial interests.
 \item[Correspondence] Correspondence and requests for materials
 should be addressed to SC~(email: sayan@tifr.res.in).
\end{addendum}

\begin{figure}
\includegraphics[width=0.85\textwidth]{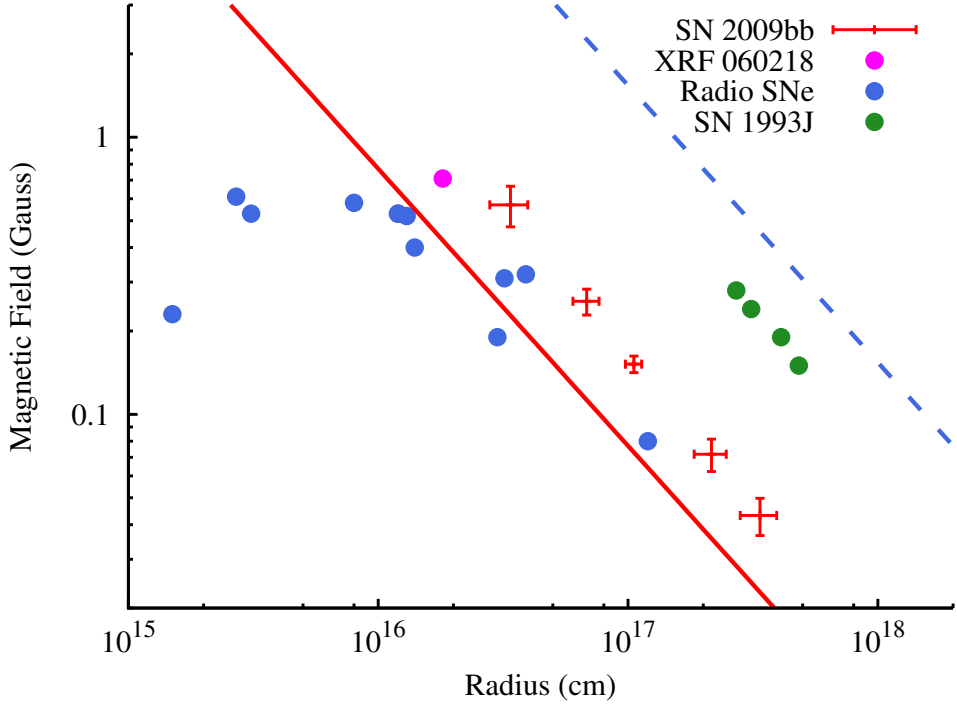}
\caption{Hillas Diagram:
Mildly relativistic sources ($\beta/\Gamma\sim1$) must lie above
the \textit{solid red line}, to be able to accelerate Iron nuclei to 60 EeV
by diffusive shock acceleration\cite{1987PhR...154....1B}, according
to $E_Z\lesssim \beta eZBR / \Gamma$ \cite{2005PhST..121..147W}.
In comparison, non-relativistic SNe
($\beta/\Gamma\sim0.05$) must
lie above the \textit{dashed blue line} to reach the same energies.
Radius and magnetic field of SN 2009bb (red crosses, at 5 epochs, determined here
from radio observations with VLA and GMRT assuming equipartition) and
XRF 060218\cite{2006Natur.442.1014S} (magenta ball) lie above the solid red line.
Other\cite{1998ApJ...499..810C} radio SNe with SSA fits are plotted as
blue balls. Only the SN 1993J magnetic fields are obtained without assuming
equipartition\cite{2004ApJ...612..974C}. Note that all of the
observed non-relativistic SNe (blue balls)
including SN 1993J (green balls) lie below the dashed blue line and are unable
to produce UHECRs unlike the mildly relativistic SN 2009bb and XRF 060218 which
lie above the red line. Sizes of crosses are twice the standard errors,
sizes of balls are bigger than the standard errors.
}
\end{figure}

\newpage
\pagenumbering{roman}


\section*{\begin{center}\textbf{Supplementary Information}\end{center}}

\subsection{Radius-Magnetic Field Evolution:}
With a robust set of assumptions for the
electron energy distribution and magnetic fields\cite{1998ApJ...499..810C},
the radius of the forward shock wave at the time of the synchrotron
self-absorption peak can be written as\cite{1998ApJ...499..810C}
\begin{equation}
R\backsimeq4.0\times 10^{14}\alpha^{-1/19} \left(f\over 0.5\right)^{-1/19}
\left(F_{op}\over {\rm mJy}\right)^{9/19}  \left(D\over {\rm Mpc}\right)^{18/19}
\left(\nu\over 5 {\rm~ GHz}\right)^{-1}  {\rm~cm},
\label{Rp}
\end{equation}
where $\alpha=\epsilon_e/\epsilon_B$ is the ratio of relativistic electron
energy density to magnetic energy density, $f$ is the fraction of the
spherical volume occupied by the radio emitting region, $F_{op}$ is the
observed peak flux, and $D$ is the distance. Using the same variables, the
magnetic field is given by
\begin{equation}
B\backsimeq1.1\alpha^{-4/19} \left(f\over 0.5\right)^{-4/19}
\left(F_{op}\over {\rm mJy}\right)^{-2/19}  \left(D\over {\rm Mpc}\right)^{-4/19}
\left(\nu\over 5 {\rm~ GHz}\right)  {\rm~G}.
\label{Bp}
\end{equation}
The radio spectrum of SN 2009bb at all epochs from discovery
paper (Fig. 2 of ref\cite{2010Natur.463..513S}) and this work,
as obtained from observations using the Very Large Array (VLA) and
the Giant Metrewave Radio Telescope (GMRT), is well fit
by the SSA model, giving us a rare
opportunity to explicitly
measure the size and magnetic field of a candidate accelerator, instead of
indirect arguments connecting luminosity with the Poynting flux.

\subsection{Equipartition:}
As already stated, non-relativistic SN have sizes and magnetic
fields which are characteristically inadequate to accelerate charged
particles to the highest energies. However the inferred magnetic fields are
based mostly on equipartition arguments. The only young SN
where  the magnetic field was determined independent of  the
equipartition assumption was SN 1993J  (this was however a type IIb
SN, unlike the type Ibc's being considered here). However, the magnetic
field determined using a synchrotron cooling break was found to be $\sim9$
times larger than the equipartition value\cite{2004ApJ...612..974C}.
This only helps by
placing the SN 1993J much closer to the Hillas line (but still below it)
due to enhanced $BR$ product. The energy requirement to explain the radio
emission has a minimum\cite{1998Natur.395..663K} at the
assumed equipartition factor of $\alpha = 1$. In the absence of an independent
measurement of the magnetic field, as in SN 1993J\cite{2004ApJ...612..974C},
Very Long Base Interferometry of the outflow can constrain the deviation
from equipartition using Equation (\ref{Rp}). However, because 
of the very slow dependence of the radius on $\alpha$, even for
a slight deviation from equipartition in SN 2009bb, it can easily
inflate the energy in the radio emitting plasma. In the case of SN 2009bb,
even a more conservative
assumption of equipartition, aided by its demonstrated mildly
relativistic outflow enables it to be in a class of SN which can readily
account for possible accelerators of UHECRs. 

\subsection{Energy Budget in SN 2009bb:}
The radius evolution of SN 2009bb, as inferred from our radio observations (Table 1),
is consistent with almost free expansion. This can only
be explained if the mass of the relativistic ejecta is still much larger than
the swept up mass. The computed Lorentz factor
has barely decreased from 1.32 to 1.23 between days 20 and 222 post explosion.
Using the model for collisional slowdown of the ejecta, we modify Equation 115 of
Ref\cite{1999PhR...314..575P}, to give
\begin{equation}
\frac{m(R_2)}{m(R_1)+M_{0}}
=-(\gamma _{1}-1)^{1/2}(\gamma _{1}+1)^{1/2}
 \int_{\gamma_{1}}^{\gamma_{2} }(\gamma' -1)^{-3/2}(\gamma' +1)^{-3/2}d\gamma' ,
\label{mgamma1}
\end{equation}
where $m(R_1)$ and $m(R_2)$ are the swept up mass at the two respective epochs.
We have neglected radiative losses, as they are unlikely to be important for
protons in the time range of interest. Moreover radiative losses would
only help increase our initial energy budget.
Performing the integral numerically from $\gamma_{1}$ to $\gamma_{2}$, the
Lorentz factors at
the two epochs and substituting for $m(R)$ using the progenitor mass loss
rate\cite{2010Natur.463..513S}, we solve for the ejecta mass to get
$M_{0}\backsimeq1.4\times10^{-3}M_\odot$. Most of the mass in the relativistic
outflow is due to baryons. The energy associated
with these relativistic protons and nuclei is found to be
$E_{Baryons}\gtrsim3.3\times10^{51}$ ergs. Compared to this blast-wave calorimetric
value, the
equipartition energy in the electrons and magnetic fields determined from SSA
fit to the radio spectrum was was reported\cite{2010Natur.463..513S} to be
$E_{eq}\backsimeq1.3\times10^{49}$ ergs. This gives the electrons only a fraction
$\epsilon\equiv\frac{\epsilon_e}{\epsilon_p}\backsimeq0.002$ of the energy in the
relativistic baryons. If $E_{Baryons}$ is distributed equally
over 10 decades in energy, it can account for $\sim0.33\times10^{51}$ ergs
of energy in UHECRs per logarithmic energy interval. Given the rate of relativistic 
SNe in the local universe, this is consistent with the
volumetric energy injection rate for UHECRs.

\subsection{Rate of X-Ray transients:}
The number density of X-Ray flares associated with UHECR accelerators
has been prescribed\cite{2009JCAP...08..026W}
assuming that, the accelerated electrons have the same
initial power-law index for their energy spectrum as the protons, and
that they lose all their energy radiatively. Using the values of the
physical parameters, motivated by SN2009bb,the number density of
active X-ray flares with a luminosity $\gtrsim \nu L_\nu$ is
then given by recasting Equation 2.5 of Ref\cite{2009JCAP...08..026W}
to give
\begin{equation}\label{eq:n_phot_flares}
\dot{n}\Delta t\backsimeq3\times10^{-7}
\left(\epsilon\over 0.002\right)
\left(\Gamma_{inj}\over 10^{44} {\rm~erg~Mpc^{-3} yr^{-1}}\right)
\left(\nu L_\nu\over 10^{40} {\rm erg~s^{-1}}\right)^{-1}
{\rm Mpc}^{-3}.
\end{equation}
SN 2009bb was observed with the Chandra ACIS-S instrument, at age 31 days,
in the energy range 0.3-10 keV. It had an X-ray
luminosity\cite{2010Natur.463..513S} of $L_X=4.4\pm0.9\times10^{39}$
erg s$^{-1}$. This luminosity and the rate of the relativistic SNe,
considered in this work, together can account for the UHECR flux, if they
remain active accelerators for $\Delta t$ of order $\sim1$ year. This is
consistent with our radio observations, which confirm that the
$BR$ product remains above the threshold throughout the
$\sim1$ year of observation (See Table 1).

\subsection{Rate of radio transients:}
The required rate of radio transients, which can supply the observed 
$\Gamma_{inj}$ is given by $\dot{n}=\Gamma_{inj}/E_{SN}$. Assuming
that the electrons and magnetic fields together have a fraction $\epsilon$
of the energy of the relativistic protons (which is assumed to be divided
equally into $\sim 10$ logarithmic bins, assuming $p\approx2$ for the
protons), we compute the minimum required
rate of such transients with peak radio luminosity $L_{op}$, which remain
mildly relativistic at least until the SSA peak frequency drops to $\nu$, as
\begin{eqnarray}
\dot{n}\backsimeq 3 \times 10^{-7}
\left(\Gamma_{inj}\over 10^{44} {\rm~erg~Mpc^{-3} yr^{-1}}\right)
\left(\epsilon \over 0.002\right)
\left(L_{op}\over {10^{29}\rm~ergs/sec/Hz}\right)^{-23/19} \nonumber \\
\times \left(\nu\over 0.5 {\rm~ GHz}\right)
\left(2\over \eta^{11} (1+\eta^{-17}) \right) {\rm~Mpc^{-3}yr^{-1}}.
\label{n}
\end{eqnarray}
Here $\eta=\theta_{obs}/\theta_{eq}$ is the ratio between the observed angular
radius and the one obtained
by assuming equipartition between electrons and magnetic
fields\cite{1998Natur.395..663K}. This criterion works for an electron
energy index between $p=2$ (with the cooling break shifted below
the SSA peak) to $p=3$ (with the cooling break above the observed
radio frequencies), so as to explain the observed spectral index of
$\approx1$ in the optically thin part of the radio spectrum. Here,
Equation (\ref{n}) is the radio analogue of Equation 2.5 of
Ref\cite{2009JCAP...08..026W} (which is for X-ray transients).
Hence, the observed rate of relativistic SNe can easily explain the
energy injection rate.

\subsection{Survey Parameters:}
The prototypical mildly relativistic SN 2009bb has been
discovered\cite{2010Natur.463..513S} in a dedicated radio follow-up of
type Ibc SNe. To firmly establish the rate of occurrence of such
relativistic SNe in the local universe, a systematic large area
radio survey is required. Here we
estimate the maximum energy to which relativistic SNe can
accelerate nuclei of charge $Ze$ from $E_Z\lesssim ZeBR$ as $\beta/\Gamma\sim1$
for mildly relativistic outflows.
Substituting the expressions for the radius (Equation \ref{Rp}) and magnetic
field (Equation \ref{Bp}) we have
\begin{equation}
E_z\backsimeq6.4 \times Z\alpha^{-5/19} \left(f\over 0.5\right)^{-5/19}
\left(F_{op}\over {\rm mJy}\right)^{7/19}  \left(D\over 200{\rm~Mpc}\right)^{14/19}
{\rm~EeV},
\label{EZ}
\end{equation}
which is independent of the observed SSA peak frequency $\nu$.
Further, assuming that $R\sim \Gamma \beta ct$
and we get the time to reach the synchrotron peak is
\begin{equation}
t_{peak}\backsimeq23\times \left(1\over \Gamma \beta\right)  
\left(F_{op}\over {\rm mJy}\right)^{9/19}  \left(D\over 200 {\rm~Mpc}\right)^{18/19}
\left(\nu_{survey}\over 5 {\rm~ GHz}\right)^{-1} {\rm~days.}
\end{equation}
Hence relativistic SNe will have faster rise times than non-relativist
radio transients, allowing them to be easily identified for multi-frequency follow
up with targeted observations. These considerations determine the cadence and
sensitivity of the proposed radio survey as mentioned in the main text.

\subsection{Propagation and Survival of Nuclei:}
In the particle acceleration scheme outlined in this work, the highest energy
particles are likely to be nuclei rather than protons. This is borne out by
the latest Auger data which favors an increasing average rest mass of primary
cosmic ray particles at the highest energies\cite{2010PhRvL.104i1101A}.
As for protons, the flux of ultra high energy nuclei are also suppressed
over cosmological distances via interaction with background radiations.
CMB photons appear has high energy $\gamma$-rays when Lorentz boosted into
the rest frames of ultra high energy protons or nuclei. Protons
above $\sim60$ EeV, interact with CMB photons via the $\Delta$ resonance
($\gamma_{\rm CMB}+p\rightarrow\Delta^+\rightarrow p + \pi^0$
or $\gamma_{\rm CMB}+p\rightarrow\Delta^+\rightarrow n + \pi^+$)
and give rise to the GZK limit\cite{1966PhRvL..16..748G,1966JETPL...4...78Z}.
Similarly ultra high energy nuclei can be photo-disintegrated by Lorentz boosted
cosmic infrared background photons interacting mainly via Giant Dipole Resonances. 
The distance over which this effect suppresses
the flux of ultra high energy nuclei is a function of the nuclear species and
its energy. Detailed calculations\cite{2007APh....27..199H} using updated
photo-disintegration cross-sections indicate that the energy loss lengths
for $100$ EeV intermediate mass nuclei such as Ne, Si and Ca are around
$\sim100$ Mpc. As discussed in the main text, relativistic supernovae can
provide enough number of UHECR sources within this distance, to be consistent
with observation of independent arrival directions for the UHECRs.

\subsection{High Energy Particle Detection:}
UHECRs from the same source but with different energies will travel by different
trajectories due to deflections by magnetic fields\cite{1995PhRvL..75..386W}.
For current estimates of the
average intergalactic magnetic field, the mean delay in the arrival time of
UHECRs, when compared to photons is found\cite{2009ApJ...693..329F}
$\langle \tau_{delay} \rangle \approx 10^5$ yrs. Hence, barring chance coincidences,
detected cosmic rays will not point back to known astrophysical
transients\cite{2008JCAP...05..006K}.
However, detected UHECRs should point back (within the errors from deflection)
to the host galaxies. As type
Ibc supernovae occur mostly in HI rich spirals, the detected correlation
with HI selected galaxies\cite{2008MNRAS.390L..88G} is consistent with
our hypothesis. Similarly
direct detection of UHECRs from say SN 2009bb is unlikely. However photo-hadron
interaction between accelerated protons or nuclei and optical photons from
the underlying SN may produce pions which then decay ($\pi^+ \rightarrow e^+ + \nu_e
+ \bar{\nu_\mu} + \nu_\mu$ or $\pi^- \rightarrow e^- + \bar{\nu_e}
+ \nu_\mu + \bar{\nu_\mu}$) to give high energy
neutrinos. Neutrinos have no electric
charge, hence they are not deflected by the intergalactic magnetic fields.
They have very low rest masses compared to their very high energies and will
travel at nearly the speed of light. These neutrinos will not be coincident with
the initial core collapse as the number of accelerated charged particles
which are the source of neutrinos grows with time.
The peak of the high energy neutrino flux will approximately coincide with the
peak in bolometric luminosity (at around a week after explosion for SN 2009bb)
as the most number of photons will be available
for interaction with the accelerated hadrons. Hence, high energy neutrinos
may by found in future by neutrino detectors like the IceCube, in directional
and rough temporal coincidence with relativistic supernovae.

\subsection{Alternative Sources: AGNs}
The arrival directions of 20 of the 27 highest energy cosmic rays detected by
The Pierre Auger Cosmic Ray Observatory, based in the southern hemisphere,
were found to be within $3.2^{\circ}$
of AGNs within 75 Mpc\cite{2007Sci...318..938T}. This leads The Pierre AUGER
Collaboration to conclude that they possibly
come from either AGNs or objects with a similar spatial distribution.
Yet no significant correlation is
seen for the UHECRs detected by the HiRes stereo events
and AGNs\cite{2008APh....30..175T} in the northern hemisphere.
Even, the degree of correlation in the Auger events now appears to be
weaker\cite{2009Hague} than what was seen by the earlier data.
It has been suggested that these cosmic rays
may be accelerated in the relativistic outflows from powerful AGNs. However,
particle acceleration to such high energies ($E=10^{20}\times E_{20}$ eV) in
turbulent shocks with bulk Lorentz factor $\Gamma$ would be accompanied by a
minimum power lost to the Poynting
flux\cite{1995PhRvL..75..386W,2009ApJ...693..329F} of
\begin{equation}
L\gtrsim10^{45}\Gamma^2E^2_{20} {\rm~erg~s}^{-1}.
\label{Lmin}
\end{equation}
Continuous sources of such
luminosity would be easily detected within 200 Mpc and their absence rules out
continuous AGN jets as the sources of a significant fraction of the UHECRs.
Instead, a new class of very
intense, short-duration AGN flares were proposed as possible 
sources\cite{2009ApJ...693..329F}. However, no such flare has been
observed until now.

\subsection{Alternative Sources: Classical GRBs}
Classical GRBs have also been suggested as one of the most promising candidates for
producing the highest energy cosmic rays\cite{1995PhRvL..75..386W,1995ApJ...449L..37M},
where protons would be accelerated by the Fermi mechanism in an ultra-relativistic
outflow. For an astrophysical source driving a magnetized plasma outflow with a
characteristic magnetic field $B$, at a velocity $v=\beta c$ (bulk Lorentz factor
$\Gamma$), out to a radius $R$, the maximum energy to which a proton of charge $e$
can be accelerated by diffusive shock acceleration\cite{1987PhR...154....1B}
is given by\cite{2005PhST..121..147W}
\begin{equation}
E_p\lesssim \left(\beta eBR\over \Gamma\right).
 \label{Emax}
\end{equation}
GRBs satisfy the minimum luminosity criterion (Equation \ref{Lmin}) derived
from the Poynting flux carried out by this
outflow\cite{1995PhRvL..75..386W,2005PhST..121..147W}. However, most
classical GRBs are found at cosmological distances
with a mean redshift of 2.8 for those discovered by the
Swift\cite{2006A&A...447..897J}, hence most GRBs cannot contribute to
the flux of cosmic rays above the GZK limit. If classical GRBs are indeed the
source of the observed flux of the UHECRs, then the observed local
rate of GRBs
implies that each GRB is required to provide of orders of magnitude more
energy\cite{2009ApJ...693..329F}, than what is available from a collapsar
scenario\cite{1999ApJ...524..262M}.

\subsection{Alternative Sources: Hypernovae}
Hypernovae with a continuous ejecta profile (with
$E_k\propto(\Gamma \beta)^{-2}$) between the non-relativistic and relativistic material
have been suggested as sources of UHECRs\cite{2007PhRvD..76h3009W}.
In such a model each shell of different
velocity accelerates particles upto a different energy and adds up to a
final power law energy spectrum with slope of $\approx-1$ for $E^2 dN/dE$,
which was claimed fits the observed UHECR spectra\cite{2007PhRvD..76h3009W}.
However, the observed spectrum of UHECRs
is suppressed beyond the GZK limit\cite{2008PhRvL.100j1101A} via interaction
with the CMB photons and has to be
corrected for the propagation effects to get the original injection spectrum.
Hence Hypernovae with continuous
ejecta profiles cannot reproduce flat injection spectrum of UHECRs which
requires equal energies in each logarithmic energy bin. Galactic trans-relativistic
SNe have also been considered\cite{2008ApJ...673..928B}. However,
this requires at least one trans-relativistic SNe per Galactic
confinement time for the cosmic ray energies being considered. 
For particles with energies beyond the GZK limit, the
magnetic rigidities are so high, that their confinement time is
comparable to the light crossing time\cite{1963PhRvL..10..146L}
of $10^4$ years\cite{2008ApJ...673..928B}. There are around
$\sim100$ SNe in this time, of which say upto $10$ are SNe Ibc.
Given the fraction of Ibc SNe which have relativistic
outflows\cite{2010Natur.463..513S}, there are $\sim0.07$ trans-relativistic
SNe in this time. Clearly, if these events are galactic, the rate of
such objects is too low and a galactic origin would neither explain the
independent arrival directions\cite{2007Sci...318..938T}
nor the GZK suppression of the spectra\cite{2008PhRvL.100j1101A}.

\begin{table}
 \begin{tabular}{| l | r r r r r r r |}
\hline			
Observation	& Age   	& $F_{op}$	& $\nu_{p}$	& $R$   	& $B$	& $E_p$	& $E_{Fe}$\\
Date (2009)   	& (Days)	& (mJy)   	& (GHz)   	& ($10^{15}$cm)	& (mG)	& (EeV)	& (EeV)\\
\hline
05 April 	& 17	& $>$24.53		&\ldots		&\ldots		&\ldots		& $>$6.4	& $>$166\\ 
08 April 	& 20	& 17.87$\pm$0.95	& 7.63$\pm$0.63	& 34$\pm$3	& 570$\pm$48	& 5.7$\pm$0.1	& 148$\pm$3\\
10 May  	& 52	& 13.69$\pm$0.79	& 3.33$\pm$0.17	& 68$\pm$4	& 256$\pm$14	& 5.2$\pm$0.1	& 134$\pm$3\\
08 June 	& 81	& 10.82$\pm$0.34	& 1.93$\pm$0.07	& 106$\pm$4	& 152$\pm$5	& 4.7$\pm$0.1	& 123$\pm$1\\
10 August 	& 144	&  9.82$\pm$0.65	& 0.90$\pm$0.06	& 216$\pm$16	& 72$\pm$5	& 4.6$\pm$0.1	& 119$\pm$3\\
27 October 	& 222	&  8.35$\pm$0.59	& 0.53$\pm$0.04	& 337$\pm$28	& 43$\pm$3	& 4.3$\pm$0.1	& 112$\pm$3\\
\hline  
\end{tabular}
\caption{Radius-Magnetic Field Evolution:
Peak fluxes and peak frequencies of SN 2009bb are determined
from VLA and GMRT observations by fitting a SSA spectrum to the
observed fluxes. Fluxes until August can
be found in Supplementary Info. of Ref\cite{2010Natur.463..513S}.
The fluxes around 27 October 2009 are from new VLA ($1.6\pm0.1$ mJy at
8.46 GHz and $3.7\pm0.2$ mJy at 4.86 GHz) and GMRT observations ($4.4\pm0.3$
mJy at 1.28 GHz, $8.7\pm0.7$ at 617 MHz and $5.8\pm0.7$ mJy at 332 MHz).
Radius and magnetic fields are determined using Equations (\ref{Rp},\ref{Bp}).
Maximum energies to which protons and Iron nuclei can be accelerated
are computed using Equation (\ref{EZ}), $\pm$ are standard errors.
Note that both $E_p$ and $E_{Fe}$ decrease slowly by only $\sim24\%$ in a span
of $\sim200$ days.}
\end{table}

\end{document}